\documentclass[12pt,preprint]{aastex}

\slugcomment{submitted to ApJ}

\shorttitle{Magellanic Cloud Bump Cepheids}
\shortauthors{Keller and Wood}

\begin{document}


\title{Bump Cepheids in the Magellanic Clouds\\Metallicities, the Distances to
 the LMC and SMC, and the Pulsation-Evolution Mass Discrepancy}


\author{S.\ C.\ Keller}
\affil{RSAA, Australian National University, Canberra A.C.T.~2600, Australia}
\email{stefan@mso.anu.edu.au}

\and

\author{P.\ R.\ Wood}
\affil{RSAA, Australian National University, Canberra A.C.T.~2600, Australia}
\email{wood@mso.anu.edu.au}



\begin{abstract}

We utilize nonlinear pulsation models to reproduce the observed light and
color curves for two samples of bump Cepheid variables, 19 from the Large
Magellanic Cloud and 9 from the Small Magellanic Cloud.  This analysis
determines the fundamental parameters mass, luminosity, effective temperature,
metallicity, distance and reddening for the sample of stars.  The use of light
curve shape alone to determine metallicity is a new modelling technique
introduced here.

The metallicity, distance and reddening distributions for the two samples are
in agreement with those of similar stellar populations in the literature. The
distance modulus of the Large Magellanic Cloud is determined to be
18.54$\pm$0.018 and the distance modulus of the Small Magellanic Cloud is
determined to be 18.93$\pm$0.024.  The mean Cepheid metallicities are $Z =
0.0091\pm0.0007$ and $0.0050\pm0.0005$ for the LMC and SMC, respectively.

The masses derived from pulsation analysis are significantly less than those
predicted by stellar evolutionary models with no or mild convective core
overshoot. We show that this discrepancy can not be accounted for by
uncertainties in our input opacities or in mass-loss physics. We interpret the
observed mass discrepancy in terms of enhanced internal mixing in the vicinity
of the convective core during the main-sequence lifetime and find that the
overshoot parameter $\Lambda_{c}$ rises from 0.688$\pm$0.009H$_{\rm p}$ at
the mean LMC metallicity to 0.746$\pm$0.009H$_{\rm p}$ in the SMC.

\end{abstract}




\section{Introduction}

Wide-field photometric studies such as the OGLE and MACHO projects have left a
legacy of well-sampled highly uniform photometry for the Cepheid populations
of the Magellanic Clouds. In the present paper we compare these data to the
predictions of stellar pulsation models. A number of recent studies have shown
that nonlinear pulsation models are able to reproduce both the gross features
of Cepheids light curves such as periods and amplitudes as well as much smaller
morphological features such as resonance features seen in bump Cepheids
\citep{woo97,bon02}.

The morphology of the light curve proves highly sensitive to the physical
parameters of mass, luminosity and effective temperature. By matching the
model light and color curves to those observed we can constrain the physical
parameters of the star to greater precision than available with other
techniques \citep{kel02}. The distance and reddening to each Cepheid can also
be determined. The resulting distances offer an independent method of distance
determination to the Magellanic Clouds: one which is solely based on the
physics of stellar interiors.

Mass estimates derived from bump Cepheids have caused much debate since
\cite{sto69} first showed that pulsation masses for bump Cepheids are
significantly lower than those predicted from stellar evolution. Largely in
response to this mass discrepancy, more detailed modelling of opacities was
undertaken by the OPAL \citep{rog92} and Opacity Projects \citep{sea94}. The
resultant enhanced metal opacities resolved much of the mass discrepancy
\citep{mos92}.  However a series of studies of galactic \citep{cap05} and LMC
\citep{woo97,kel02,bon02} bump Cepheids have shown that the masses of bump
Cepheids remain smaller than predicted by evolution. This discrepancy amounts
to 15\% in mass or, seen another way, bump Cepheids appear $\sim20\%$ more
luminous for a given mass \citep{kel02}.

The luminosity of a Cepheid is critically dependent on the mass of the He core
established during the star's main-sequence life. The mass of the He-core is
dependent on the extent of the convective core during core-H
burning. Classical models define the limit to convection via the Schwarzschild
criterion. This places the boundary to convection at the radius at which the
buoyant force drops to zero. However, the temperature and density regime in
the vicinity of the convective boundary of the main-sequence Cepheid
progenitor are such that restorative forces in the region formally stable to
convection are mild and some significant level of overshoot of the classical
boundary is expected \citep{zahn91}.

It is typical to parameterize the level of convective core overshoot using the
mixing-length formulation where gas packets progress a distance given by
$\Lambda_{c}$ pressure-scale heights ($H_{\rm p}$) into the classically stable
region. A star with larger core overshoot is able to draw upon more H and
hence lives longer on the main-sequence, develops a more massive He core and
is more luminous than classical models during the post-main-sequence
evolution.

Our description of convection is the weakest point in our understanding of the
physics of massive stars. Numerical modelling of core convection requires a
description of the turbulence field at all scales. Three-dimensional
hydrodynamical calculations capable of adequate resolution have only recently
become a possibility and are in their infancy \citep{djehuty}. At present we
must rely on observation for constraint.

Observational determinations of the magnitude of convective core overshoot
(CCO) have focused on populations of intermediate mass stars ($M=5-12$
M$_{\odot}$) where the signature of overshoot is expected to be most clearly
seen. The study of \cite{mer86} examined a range of galactic clusters with
main-sequence turn-offs in mass range 9-15M$_{\odot}$ and derived overshoot
parameter $\Lambda_{c} \sim 0.6$$H_{\rm p}$.\footnote{We quantify CCO using the
formalism of \cite{bre81} who use an overshoot parameter $\Lambda_{c}$
pressure scale heights.  Note that $\Lambda_{c}$ is a factor of 2.0 times the
overshoot parameter $d_{\rm over}/H_{\rm p}$ in the formalism of the Geneva
group \citep{schaller92}.} \cite{chiosiARAA92} summarize a number of similar
studies converging on a mild efficiency of overshoot $\Lambda_{c} \sim
0.5$. This is the basis for the level of CCO included in the stellar
evolutionary models for $M > 1.3$M$_{\odot}$ by \cite{gir00}.

A number of subsequent studies have focused on the intermediate age LMC
cluster NGC 1866 whose large population of post-main-sequence stars was
proposed as proof both for \citep{bar02} and against \citep{testa99} mild
overshoot. \cite{kel01} examined the populations of a series of young populous
clusters in the Magellanic Clouds and found $\Lambda_{c}=0.67\pm0.20$$H_{\rm
p}$ for the SMC cluster NGC 330 and mean of $\Lambda_{c}=0.61\pm0.12$$H_{\rm
p}$ for three similar clusters in the LMC. This result does not indicate any
strong dependence of $\Lambda_{c}$ on metallicity. The subsequent study of
\cite{cor02} examined the field population of the Magellanic Clouds and in
contrast found evidence for a strong metallicity dependence in the degree of
overshoot ($\Lambda_{c}=0.8$$H_{\rm p}$ and 0.2$H_{\rm p}$ for the SMC and LMC
respectively. Such studies based on the color-magnitude diagrams of stellar
populations do face uncertainty due to contamination by unresolved binary
systems (and significant blending of unrelated stars in the case of the
Magellanic Clouds), contamination by the surrounding field population, small
numbers of stars on the upper main-sequence and systematics introduced by
bolometric corrections to place the population on the H-R diagram.
 
\cite{bon02} reported the results of light curve modelling of two LMC bump
Cepheids and found that they were able to match the luminosities of the
Cepheids using masses 15\% less than predicted by evolutionary models which
neglect convective core overshoot and mass-loss. \cite{kel02} (Paper I)
presented results of pulsation modelling for a sample of 20 LMC bump
Cepheids. Paper I found a level of core overshoot of
$\Lambda_{c}=0.65\pm0.03$$H_{\rm p}$ under the assumption of a LMC abundance
of Z=0.008. The models of Paper I incorporated convective energy transfer in
the Cepheid envelope through the mixing-length approximation. This
approximation is expected to break down at cooler temperatures as the
convective region of the Cepheid becomes a substantial fraction of the
envelope. A consequence is that these models do not reproduce the red edge of
the instability strip. Redward of the red edge the pulsation growth rate
remains above zero. Consequently in Paper I we were restricted to bump
Cepheids close to the blue edge of the instability strip.  In addition,
\cite{feu00} demonstrates that neglect of the effects of convection can shift
the blue edge to temperatures on the order of 400K cooler than predicted by
models incorporating the effects of turbulent convection.

In this paper, we present updated models which include convection and
turbulent eddy viscosity. The formulation of our models is discussed in
Section 2. In Section 3 we analyze for the first time a sample 9 SMC stars and
reanalyse 19 LMC Cepheids for comparison. We then discuss the implication of
our results for the level of convective core overshoot and its dependence on
metallicity in Section 4.

\section{Model details}

The details of the nonlinear pulsation code have been presented in
\cite{woo74} and \cite{woo97}. Briefly, opacities are drawn from OPAL 96
(Iglesias \& Rogers 1996) and supplemented at low temperatures by Alexander \&
Ferguson (1994). A new addition to the pulsation code is eddy viscosity
pressure $P_{\nu}$, implemented using the prescription given by \cite{kol02}.
Given the convective velocity $v_{\rm c}$ from the time-dependent mixing
length theory described by \cite{woo74}, $P_{\nu}$ is given by
$P_{\nu}=\sqrt{\frac{2}{3}}\alpha_{\nu}\Lambda\rho v_{\rm c}r\frac{\partial
~}{\partial r}(\frac{v}{r})$, where $\Lambda$ is the mixing length, $v$ is the
pulsation velocity and $\alpha_{\nu}$ is a dimensionless parameter for which
we use a value 0.15, similar to the value 0.164 used by \cite{kol02}.

Envelope models were defined using $\sim$460 radial points outside an radius
of 0.3 R$_{\odot}$. A linear non-adiabatic code was used to derive the
starting envelope model. This model was perturbed by the fundamental mode
eigenfunction and the pulsation followed in time with the nonlinear code until
the kinetic energy of pulsation reached a steady limit cycle (typically
requiring around 500 cycles).

\section{Light curve modelling}

Photometry for our sample of bump Cepheids is taken from the MACHO photometric
database. The MACHO $B$ and $R$ magnitudes have been converted to Kron-Cousins
V and R using the transformations described in Alcock et al. (1999). Alcock et
al. quote global systematic uncertainties of $\pm0.035$ mag in zero point and
$V$$-$$R$ color.
The photometric parameters of our Cepheid sample are presented in Table 1.

The appearance of a bump in the light curve of Cepheids of periods
6d$<$P$<$16d and its dependence on period was first described by
\cite{hea26}. This Hertzsprung progression starts at short periods with a
small amplitude bump on the descending branch of the light curve. As the
period increases the amplitude increases and the phase of the bump decreases
until the bump appears at maximum light. This is the centre of the Hertzsprung
progression. With increasing period the bump amplitude decreases and it moves
along down the rising branch of the light curve. From a Galactic sample the
center of the Hertzsprung progression is seen to occur at P=9.95$\pm$0.05 d
\citep{mos00}. The study of \cite{bea98} finds that in the LMC this occurs at
P=10.5$\pm$0.5d and in the SMC at P=11.0$\pm$0.5 d. The theoretical
investigation of the Hertzsprung progression by \cite{bon00} determined the
center of the Hertzsprung progression in the LMC to be at $11.2\pm0.5$ d, in
good agreement with the empirial evidence.
The shape of a Cepheid light curve depends upon the mass, luminosity and
effective temperature of the star together with its metallicity. Hence four
constraints are required to fully define the parameters of our chosen
star. The first constraint is that the fundamental frequency of pulsation must
match that observed. The remaining three constraints are derived from fitting
the nonlinear pulsation model light curves to the observed light curve. Our
nonlinear models span a three dimensional parameter space defined by:
effective temperature, metallicity and P02, the ratio of the fundamental to
the second overtone period. This choice of parameters is driven by the
simplicity they bring to subsequent analysis, as each parameter acts on a
separate feature of the light curve. The effective temperature determines the
amplitude of pulsation. The quantity P02 determines the phase of the bump and
the metallicity affects the phase of a tertiary bump on the declining side of
the light curve.

Determining an optimal match to the observed light curve involved a chi-squared
minimization in this parameter space. For trial values of effective
temperature and P02 we determine a mass and luminosity to give the observed
period of pulsation. We then examine the light curve from the trial and form a
chi-squared figure of merit relative to the observed light curve. This is
repeated over a grid of metallicity values of $\Delta$Z=0.001 step size.  

Figures 1 and 2 show the effects of variation of the parameters $T_{eff}$, P02
and Z. As we increase temperature in the vertical direction on Figure 1 the
amplitude of pulsation is decreased as we approach the blue edge of the
instability strip. As we vary P02 we modify the phase at which the bump is
located.

In Figure \ref{figstar2}, we see that the effect of changing Z is to modify
the phase of a third bump evident during the decline in light: this bump is
also evident in the extremely well sampled MACHO light curves.  We note that
the metallicity bump is not one which has been discussed previously in the
literature, perhaps not surprisingly given the low amplitude of the
feature. As metallicity is decreased the feature descends the declining branch
of the light curve and the amplitude diminishes. The feature is no longer
apparent by Z$\sim$0.002 since it has moved into the upturn region of the
light curve.  Given this situation, it seems that this method of determining
metallicity is useful from solar to one tenth solar.

Having located the optimal match to the observed light curve the observed V-R
color curve is then dereddened to the model V-R color curve to derive the
color excess E$_{V-R}$. This can be converted to a visual absorption via the
standard reddening curve (\cite{bes98} based on \cite{mat90}). Likewise, the
apparent distance modulus to the star is simply the magnitude difference
between the model absolute magnitude $M_{V}$ and the observed mean $V$
magnitude. The apparent distance modulus minus the visual absorption provides
the true distance modulus. The parameters of the best-fit models for the
Cepheids of our sample are given in Table 2.

The internal accuracy of the best model parameters is limited by the presence
of cycle-to-cycle variations in the pulsation model output. These variations
arise since the models have not reached a perfectly steady pulsation limit
cycle. In practice the approach to the limit cycle becomes prohibitively slow
and one must accept some level of inter-cycle variation. These variations
typically amount to around $\sigma_{V}=0.03$ magnitudes.
The inter-cycle variations place a strong limit on our ability to determine
the metallicity, with typical uncertainties per star ranging from $\pm0.001$
to 0.004 in the value of Z.

The uncertainty in MACHO photometric calibrations noted above lead to
systematic uncertainties.  The uncertainties in mass and bolometric luminosity
for each star from this source are $\pm0.1$ M$_{\odot}$ and $\pm0.014$ dex in
$\log L$, respectively. The values presented in figures \ref{delmfig} and
\ref{delm_zfig} are the sum in quadrature of systematic and internal
uncertainties.

\subsection{Reddenings and distance modulii}

In Figures \ref{redfig} and \ref{dmfig} we present the distributions of
derived reddenings and distance modulii for our LMC and SMC samples. The mean
derived reddening is 0.08$\pm$0.02 for the LMC and 0.12$\pm$0.02 for the
SMC. \cite{bes91} summarizes a large body of previous reddening investigations
for the LMC and SMC and concludes that average reddening of stars in both
systems is of order $E(B-V)$=0.10 but importantly, different populations give
systematically different results. The studies of \cite{zar99} and \cite{zar02}
describes this systematic effect in more detail. In the LMC \cite{zar99} find
a mean $E(B-V)$=0.03 from a red-clump giant sample and $E(B-V)$=0.14 from OB
stars. Both populations exhibit non-Gaussian tails extending to
$E(B-V)$$\sim$0.3. The results are similar for the SMC: $E(B-V)$=0.06 from
red-clump giants and $E(B-V)$=0.14 from OB types \citep{zar02}. \cite{zar02}
propose that the variation in reddening between populations is due to an age
dependent scale height. OB stars having a smaller scale height lie
predominantly in the dusty disk. With this background, we find that the
reddenings derived for our sample are applicable to the range of values
expected in the two localities.

The mean distance modulus is 18.54$\pm$0.018 for the LMC and 18.93$\pm$0.024
for the SMC. Both are in good agreement with existing determinations in the
literature. In the case of the LMC, determinations of the distance modulus
from population I and II distance indicators cover the range of 18.3-18.7
magnitudes. Since 2002 there has been a convergence towards a value of
18.50$\pm$0.02mag \citep{alv04}.  A recent pulsation-based determination of
LMC distance modulus using light curve fitting to RR Lyrae stars rather than
Cepheids \citep{mar05} gave a value of $18.54\pm0.02$, essentially identical
to the value found here.  The distance to the SMC is much less well defined,
probably due to line-of-sight depth effects within the galaxy. Using stellar
population studies, \cite{uda99} and \cite{gro00} have estimated that the
differential distance modulus of the SMC from the LMC is +0.5mag. \cite{hil05}
used a sample of 40 eclipsing binaries in the SMC to derive a mean distance
modulus to the SMC of 18.91$\pm$0.03 magnitudes ($\pm0.1$mag systematic).


\subsection{Metallicity distribution for Cepheids in the Magellanic Clouds}

As we have seen from the above light curve modelling it is possible to
estimate the metallicity of the target bump Cepheid from the phase of the
secondary bump. The distribution of metallicities so derived for the LMC and
SMC samples is shown in Figure \ref{zfig}. The mean metallicity is
0.0091$\pm$0.0007 for the LMC ([Fe/H] = $-0.34\pm0.03$) and 0.0050$\pm$0.0005
for the SMC ([Fe/H] = $-0.64\pm0.04$). This result is in line with previous
determinations of metal abundance within the young population of the
Magellanic Clouds. The recent study of the metal abundance of a sample of 12
LMC and 12 SMC Cepheids by \cite{rom05} reveals a mean [Fe/H]=-0.4 and -0.7
respectively with associated rms of $\sim$ 0.15 mag. \cite{luc98} present
abundances for a sample of 6 SMC and 10 LMC Cepheids. In the LMC they derive a
mean [Fe/H]=-0.30 with stars ranging from [Fe/H]=-0.55 to -0.19, while in the
SMC they derive a mean [Fe/H]=-0.68 with a range of [Fe/H] from -0.84 to
-0.65.

\section{The Cepheid pulsation and evolution mass discrepancy}

Our pulsation modelling has provided us with stellar masses for a set of bump
Cepheids. We can now compare these pulsation masses with masses predicted by
evolutionary models. From an evolutionary perspective, Cepheids are understood
to be crossing the instability strip along the so-called blue loops following
the initiation of core-He burning. The luminosity of these loops is relatively
constant during the passage through the instability strip. The loops therefore
define an evolutionary mass-luminosity relationship for Cepheids.

The mass-luminosity relation is a function of metallicity. Less metal-rich
Cepheids appear more luminous. In order to derive the evolutionary mass of
each of our Cepheids, we use the evolutionary tracks of \cite{gir00}, which
are defined for a wide range of metallicities.  We interpolate between the
tracks to the model-derived metallicity of each Cepheid. The known luminosity
then gives the corresponding evolutionary mass.

In Figure \ref{delmfig}, we plot the difference between the pulsation mass,
$M_{P}$, and classical evolutionary mass, $M_{E,0}$, corresponding to models
without convective core overshoot on the main-sequence i.e. $\Lambda_{\rm c} =
0$.  We also show the effects of convective core overshoot on this diagram by
displaying the difference between $M_{P}$ and evolutionary mass for models
with $\Lambda_{\rm c}=0.5$ and 1.0H$_{\rm p}$ (Fagotto et al. 1994 and
A.~G.~Bressan 2001, private communication). The figure clearly shows that the
pulsation masses are significantly smaller than those predicted by classical
evolutionary models, and they are also smaller than the commonly implemented
``mild'' convective core overshoot models which have $\Lambda_{c}$ =
0.5H$_{\rm p}$.

It has been suggested by \cite{cap05} and \cite{bon02} that mass loss is
responsible for the reduction of mass. As opposed to the models of Caputo et
al.\ and Bono, Castellani, \& Marconi that neglect mass loss, the models of
\cite{gir00} that we show in Figure \ref{delmfig} do incorporate mass-loss. In
the mass regime of interest to the present work, it is important to note that
the observational study of \cite{dej88} shows negligible mass loss prior to
the RGB phase and similarly on the blue loop excursion.

It is well known that significant mass loss can occur during the RGB phase.
In the models of Girardi et al. this is modeled using the parameterized
empirical fit $dM/dt = -4\times10^{-13}\eta L/gR$ \citep{rei75} where
$\eta=0.4$. The value of $eta$ is derived from consideration of the mass of
stars on the horizontal branch (HB), however a single value of $eta$ can not
account for the range of masses seen on the HB. Typical masses range from
0.8M$_{\odot}$ for red HB stars to $<$0.5M$_{\odot}$ for extreme HB stars. This
is a long standing problem: \cite{yon00} suggests that mass loss rates of
$10^{-9}$ to $10^{-10}$ M$_{\odot}/yr$ can produce a population of extreme HB
stars. \cite{vin02}, on the other hand, compute mass loss rates from radiation
pressure and find they are insufficent to give rise to mass loss of this order
on the HB. 

If we consider the distribution of effective temperatures for HB stars is
entirely due to variable mass loss on the RGB we must allow $\eta$ to range
from 0 to more than 0.4. The upper range of $eta$ is not easily defined as
there exists no detailed modern investigation of the effects of varying $eta$
beyond the recent evolutionary models of \cite{pie04} who consider $eta$=0.2
and 0.4. To account for the mass discrepancy for a 5M$_{\odot}$ star would
require mass loss of some 0.8M$_{\odot}$ during the RGB phase. By contrast,
the models of \cite{gir00} loose $\sim$0.03M$_{\odot}$. To account for the
observed mass discrepancy $\eta$ would have to be some $\sim$20-30 times
larger which hardly seems plausible.  We therefore conclude that it is
unlikely that mass loss on the RGB can explain the observed Cepheid mass
discrepancy.

Attempts to measure the mass loss rate during the Cepheid phase have been made
by \cite{dea88} who utilized the infrared excess from IRAS data in combination
with UV line profiles from IUE spectra. Mass loss rates spanned the range from
$10^{-6}$ to $10^{-10}$ M$_{\odot}/yr$. However, the high end of mass loss
range is defined by only one object in the sample of Deasy namely RS Pup,
well-known for its surrounding nebulosity. The majority of Cepheids are
characterised by mass loss rates of the order of $10^{-8}$
M$_{\odot}/yr$. Furthermore, mass loss rates for Cepheids are not found to be
significantly different from those of non-variable supergiants in the vicinity
of the instability strip. \cite{wel88} place upper limits on the mass loss
rate from radio observations of $<10^{-7}$ M$_{\odot}/yr$. Both studies
conclude that mass loss is insufficent in and of itself to resolve the Cepheid
mass discrepancy.

Uncertainty in opacity is also unlikely to explain the mass discrepancy (via a
change in derived pulsation mass). Critical to the pulsation properties of
Cepheids is the Z-bump opacity arising from the dense spectrum of transitions
originating from highly ionized Fe.  Inclusion of these transitions in the
works of OPAL \citep{rog92} and OP \citep{sea94} resulted in a substantial
increase in opacity at $\log T \simeq 5.2$. The Opacity Project \citep{bad05}
have recently included further details of atomic structure (in particular, the
treatment of atomic inner-shell processes) in their calculation of opacity.
The new opacities of \cite{bad05} do show an increase over the 1992 OP and
OPAL values of opacity in the Z-bump but at a level of only 5-10\%. To account
for the mass discrepancy we observe, the opacity would need to be raised by
40-50\%, equivalent to the increase between the early Los Alamos opacities
\citep{coxtabor1976} and 1992 OPAL-OP opacities.

In Figure \ref{delm_zfig} we examine mass discrepancy as a function of
metallicity.  There is a clear increase in the fractional mass discrepancy
with decreasing metallicity: the slope of the correlation seen in Figure
\ref{delm_zfig} is different from zero by more than 5-sigma. In order to
produce the observed change in mass discrepancy with metallicity, it is
necessary for $\Lambda_{c}$ to increase with decreasing Z.  The changes in
mass discrepancy are small, rising from $\Delta M/M_{E,0}$=
0.1692$\pm$0.0022\footnote{Quoted uncertainties are 1$\sigma$ from Monte-Carlo
realizations of the data using the individual quoted uncertainties}
($\Lambda_{c}=0.688\pm0.009 H_{p}$) at the derived mean LMC metallicity of
Z=0.0091 to 0.1836$\pm$0.0021 ($\Lambda_{c}=0.746\pm0.009 H_{p}$) at SMC
metallicity of Z=0.0050.

In showing that the extent of the convective core is significantly larger than
that defined by the Schwarzschild criterion, we can not determine its physical
origin. In addition to the ad-hoc convective core overshoot model we have
discussed above, one natural way to bring about larger internal mixing is via
the effects of rotation. The sheer layer formed at the interface between
convective and radiative regions has been shown to lead to larger He core size
\citep{heg00,mey00}. Furthermore, rotation can account for the range in
surface abundance modifications seen in stars of $M>5$M$_{\odot}$ during both
the pre- and post-first-dredge-up stages \citep{ven95,duf00,daf01,gie92}.
Furthermore, \cite{ven99} finds evidence for a range of N abundance
enhancements amongst a sample of SMC A supergiants that is much greater than
that found in Galactic counterparts. Taken together, these various
observations suggest that average stellar rotation increases in lower
metallicity populations (or at least that it increases from the Galaxy to the
LMC to the SMC).  Such an effect has been seen directly by \cite{kel04} who
made a comparison of the rotation velocity distribution of M$=5-12$M$_{\odot}$
main-sequence stars in the LMC and the Galaxy. The findings of the present
paper qualitatively match that expected from the rotationally-induced mixing
paradigm provided rotation increases with decreasing metallicity.

\section{Conclusions}

By matching the observed light and color curves of 19 LMC and 9 SMC bump
Cepheids we have been able to place tight constraints on the fundamental
stellar parameters of mass, luminosity, effective temperature and metallicity
and the secondary parameters of distance and reddening.  The use of light
curve fitting to derive metallicity is a new technique developed here.  We
have revisited the mass discrepancy that exists between the masses derived
from pulsation modelling and the masses predicted from stellar evolution for
bump Cepheids. The derived pulsational masses are significantly less than
expected from stellar evolution models that do not incorporate extension to
the convective core during main-sequence evolution.  Significant overshoot of
the convective core during the main-sequence phase is required to bring
pulsation and evolutionary masses into agreement.  In addition, we find that
the amount of overshoot is a function of metallicity.  The level of convective
core overshoot rises from 0.688$\pm$0.009H$_{\rm p}$ at LMC metallicity to
0.746$\pm$0.009H$_{\rm p}$ at typical SMC metallicity. This trend with
metallicity is qualitatively in line with expectation from models that
incorporate rotationally-induced mixing provided rotation increases with
decreasing metallicity.

Work is currently underway to further refine our technique by the comparison
of radial velocity measurements with those predicted by our models. This
offers a more stringent test of the details of our pulsation models without
limitations imposed by systematic photometric uncertainties (viz.\ bolometric
corrections, photometric transformations and zeropoints).

\acknowledgments 

We thank A.\ Bressan et al.\ for providing us with unpublished evolutionary
models for $\Lambda_{c}=1.0$H$_{\rm p}$. This paper utilizes public domain
data obtained by the MACHO Project, jointly funded by the US Department of
Energy through the University of California, Lawrence Livermore National
Laboratory under contract No. W-7405-Eng-48, by the National Science
Foundation through the Center for Particle Astrophysics of the University of
California under cooperative agreement AST-8809616, and by the Mount Stromlo
and Siding Spring Observatory, part of the Australian National University.

\clearpage
 
\begin{table}
\begin{center}
\caption{The selected MACHO bump Cepheid sample\label{tbl-1}}
\begin{tabular}{lcccccl}
\tableline\tableline
{\it{MACHO}} star id & RA (J2000) & Dec (J2000) & $<$$V$$>$ & $<$$V$-$R$$>$ & $P$[d] & Other ID\\
\tableline
212.16079.23 & 00 51 50.4 & -73 02 30 & 14.53 & 0.54 & 9.7300 &    \\
206.17173.5 & 01 09 04.8 & -72 20 15 & 14.41 & 0.56 & 9.1585 & HV 2087  \\
212.15697.3 & 00 45 43.8 & -73 23 54 & 14.63 & 0.66 & 8.8492 &    \\
207.16317.9 & 00 55 52.0 & -72 22 35 & 14.59 & 0.43 & 7.7715 & HV 1666  \\
212.16193.25 & 00 53 57.4 & -73 01 15 & 14.72 & 0.47 & 7.4978 & HV 1599  \\
212.16024.17 & 00 51 24.7 & -72 56 43 & 14.53 & 0.38 & 7.2283 & HV 1527  \\
212.16079.42 & 00 51 38.1 & -73 01 43 & 14.99 & 0.46 & 6.6606 &    \\
212.16015.8 & 00 50 47.0 & -73 02 30 & 14.99 & 0.57 & 6.5466 & HV 1512  \\
211.16704.8 & 01 01 38.9 & -73 10 23 & 15.18 & 0.46 & 6.4911 &    \\
79.4657.3939 & 05 08 49.4 & -68 59 59 & 14.23 & 0.43 & 13.8793 &    \\
9.4636.3 & 05 09 04.5 & -70 21 55 & 14.16 & 0.41 & 13.6315 &    \\
2.4661.3597 & 05 09 16.0 & -68 44 30 & 14.32 & 0.39 & 11.8591 & HV 905  \\
1.3692.17 & 05 02 51.4 & -68 47 06 & 14.53 & 0.39 & 10.8552 &    \\
1.3441.15 & 05 01 52.0 & -69 23 23 & 14.45 & 0.37 & 10.4136 & HV 2277  \\
1.3812.15 & 05 03 57.3 & -68 50 25 & 14.61 & 0.39 & 9.7118 & OGLE LMC-SC14 178619 \\
18.2842.11 & 04 57 50.2 & -68 59 23 & 14.83 & 0.38 & 8.8311 &    \\
79.5139.13 & 05 11 53.1 & -69 06 49 & 14.59 & 0.38 & 8.7716 &    \\
19.4303.317 & 05 06 39.9 & -68 25 13 & 14.65 & 0.37 & 8.7133 & OGLE LMC-SC13 194103 \\
79.5143.16 & 05 12 18.8 & -68 52 46 & 14.61 & 0.34 & 8.2105 &    \\
79.4778.9 & 05 09 56.3 & -68 59 41 & 14.56 & 0.33 & 8.1868 &    \\
77.7189.11 & 05 24 33.3 & -69 36 41 & 14.73 & 0.39 & 7.7712 &    \\
1.4048.6 & 05 05 08.8 & -69 15 12 & 14.77 & 0.36 & 7.7070 &    \\
77.7670.919 & 05 27 55 & -69 48 05 & 14.85 & 0.34 & 7.4423 &    \\
9.5240.10 & 05 13 10.1 & -70 26 47 & 15.11 & 0.38 & 7.3695 & HV 2386  \\
9.5608.11 & 05 15 04.7 & -70 07 11 & 14.81 & 0.35 & 7.0693 & HV 919  \\
78.6581.13 & 05 20 56.0 & -69 48 19 & 14.97 & 0.36 & 6.9306 &    \\
19.4792.10 & 05 09 36.9 & -68 02 44 & 14.96 & 0.36 & 6.8628 & HV 2337  \\
6.6456.4346 & 05 20 23.1 & -70 02 33 & 15.16 & 0.36 & 6.4816 &    \\
\tableline
\end{tabular}
\end{center}
\end{table}

\begin{table}
\begin{center}
\caption{The derived properties of the bump Cepheids\label{tbl-2}}
\begin{tabular}{lrccccccc}
\tableline\tableline
{\it{MACHO}} star id & $P$[d] & $P_{02}$ & log($T_{\rm eff}$) & E($B$-$V$) & $\mu$ & log($L$/L$_{\odot}$) & $M$/M$_{\odot}$ & $Z$\\
\tableline
212.16079.23 & 9.7300 & 1.980 & 3.763 & 0.138 & 18.940 & 3.61 & 5.57 & 0.005 \\
206.17173.5 & 9.1585 & 1.985 & 3.765 & 0.112 & 18.895 & 3.58 & 5.15 & 0.004 \\
212.15697.3 & 8.8492 & 1.970 & 3.760 & 0.155 & 18.893 & 3.52 & 5.08 & 0.006 \\
207.16317.9 & 7.7715 & 1.930 & 3.763 & 0.077 & 19.088 & 3.48 & 4.89 & 0.004 \\
212.16193.25 & 7.4978 & 1.950 & 3.765 & 0.127 & 18.873 & 3.46 & 4.79 & 0.004 \\
212.16024.17 & 7.2283 & 1.945 & 3.767 & 0.104 & 18.989 & 3.68 & 5.44 & 0.008 \\
212.16079.42 & 6.6606 & 1.945 & 3.772 & 0.120 & 18.901 & 3.33 & 4.42 & 0.006 \\
212.16015.8 & 6.5466 & 1.920 & 3.768 & 0.131 & 18.849 & 3.40 & 4.71 & 0.004 \\
211.16704.8 & 6.4911 & 1.940 & 3.767 & 0.123 & 18.933 & 3.34 & 4.11 & 0.004 \\
79.4657.3939 & 13.8793 & 2.035 & 3.755 & 0.104 & 18.499 & 3.82 & 6.57 & 0.017 \\
9.4636.3 & 13.6315 & 2.037 & 3.755 & 0.074 & 18.530 & 3.85 & 6.51 & 0.008 \\
2.4661.3597 & 11.8591 & 2.022 & 3.760 & 0.061 & 18.544 & 3.66 & 5.66 & 0.008 \\
1.3692.17 & 10.8552 & 1.980 & 3.767 & 0.068 & 18.515 & 3.66 & 5.86 & 0.006 \\
1.3441.15 & 10.4136 & 1.980 & 3.767 & 0.073 & 18.555 & 3.64 & 5.69 & 0.004 \\
1.3812.15 & 9.7118 & 1.968 & 3.766 & 0.079 & 18.478 & 3.64 & 5.79 & 0.015 \\
18.2842.11 & 8.8311 & 1.954 & 3.767 & 0.049 & 18.407 & 3.57 & 5.32 & 0.006 \\
79.5139.13 & 8.7716 & 1.948 & 3.768 & 0.072 & 18.495 & 3.57 & 5.42 & 0.006 \\
19.4303.317 & 8.7133 & 1.950 & 3.765 & 0.047 & 18.396 & 3.60 & 5.66 & 0.009 \\
79.5143.16 & 8.2105 & 1.952 & 3.763 & 0.119 & 18.546 & 3.55 & 5.46 & 0.010 \\
79.4778.9 & 8.1868 & 1.951 & 3.762 & 0.072 & 18.601 & 3.51 & 5.36 & 0.008 \\
77.7189.11 & 7.7712 & 1.940 & 3.763 & 0.092 & 18.613 & 3.51 & 5.36 & 0.010 \\
1.4048.6 & 7.7070 & 1.935 & 3.767 & 0.064 & 18.604 & 3.44 & 5.09 & 0.008 \\
77.7670.919 & 7.4423 & 1.932 & 3.767 & 0.073 & 18.584 & 3.50 & 5.24 & 0.008 \\
9.5240.10 & 7.3695 & 1.956 & 3.767 & 0.088 & 18.673 & 3.31 & 4.75 & 0.010 \\
9.5608.11 & 7.0693 & 1.935 & 3.760 & 0.117 & 18.655 & 3.46 & 5.24 & 0.015 \\
78.6581.13 & 6.9306 & 1.938 & 3.762 & 0.088 & 18.587 & 3.40 & 4.93 & 0.009 \\
19.4792.10 & 6.8628 & 1.905 & 3.772 & 0.051 & 18.423 & 3.43 & 4.89 & 0.006 \\
6.6456.4346 & 6.4816 & 1.929 & 3.776 & 0.093 & 18.624 & 3.22 & 4.46 & 0.008 \\
\tableline
\end{tabular}
\end{center}
\end{table}

\clearpage
\begin{figure}
\plotone{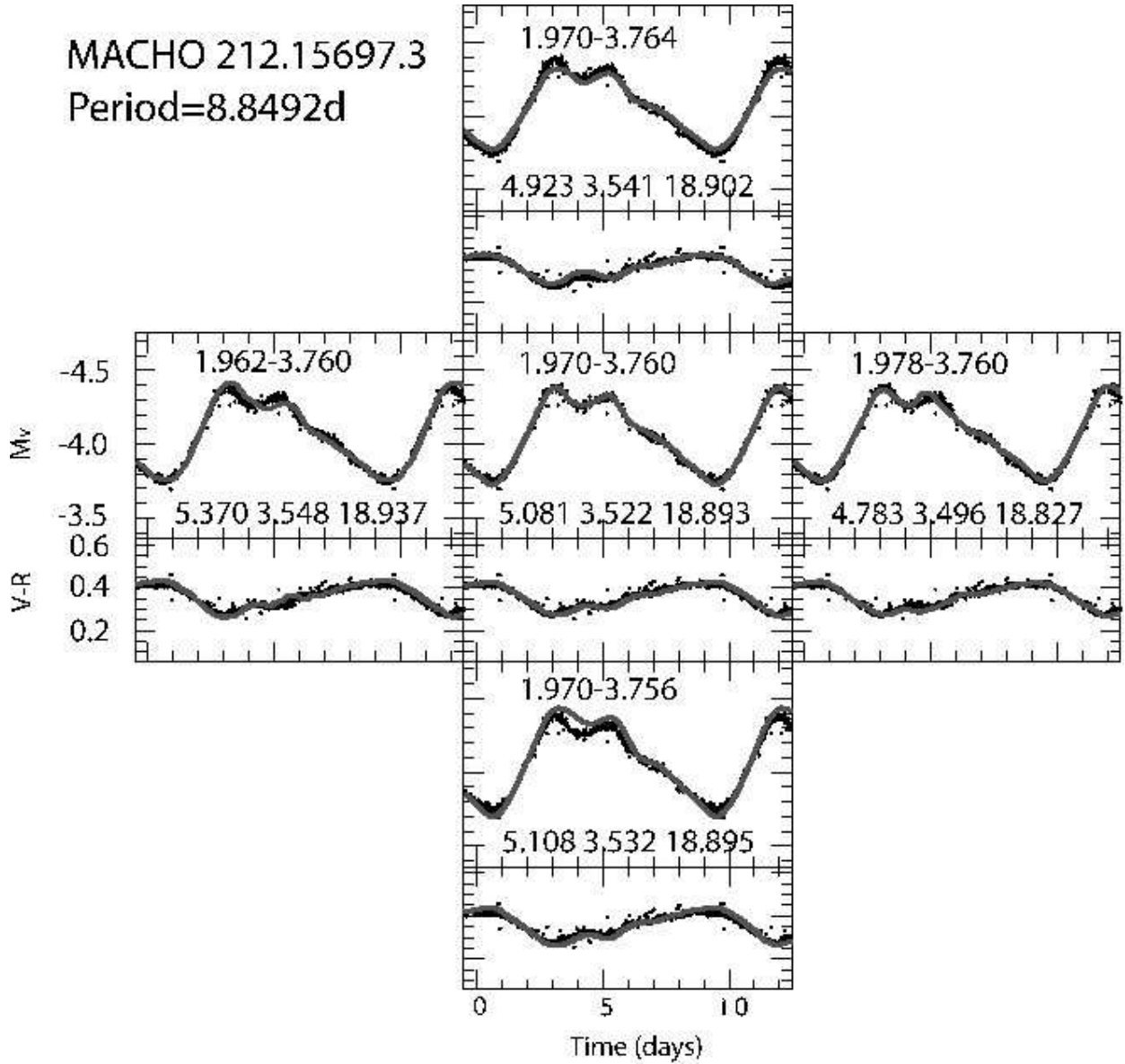}
\caption{Model fits for MACHO~212.15697.3. The numbers in each panel are: the
  period ratio P$_{02}$ and log T$_{eff}$ in the upper part, and
  M/M$_{\odot}$, log L/L$_{\odot}$, and the true distance modulus in the lower
  part. Each panel contains light and color curves (dereddened). Observations
  are shown as dots and models as lines. The panels in the vertical section
  show the effect of changing T$_{eff}$, this affects the amplitude (in the
  upper panel T$_{eff}$ is too large, the resulting amplitude is too low). The
  horizontal section shows the effect of a changing P$_{02}$ this affects the
  phase of the bump (in the left panel P$_{02}$ is too small, the phase of the
  bump is too ``late''). The central panel is the best model. All models
  incorporate a metallicity of Z=0.006.}\label{fig5panel}
\end{figure}

\begin{figure}
\plotone{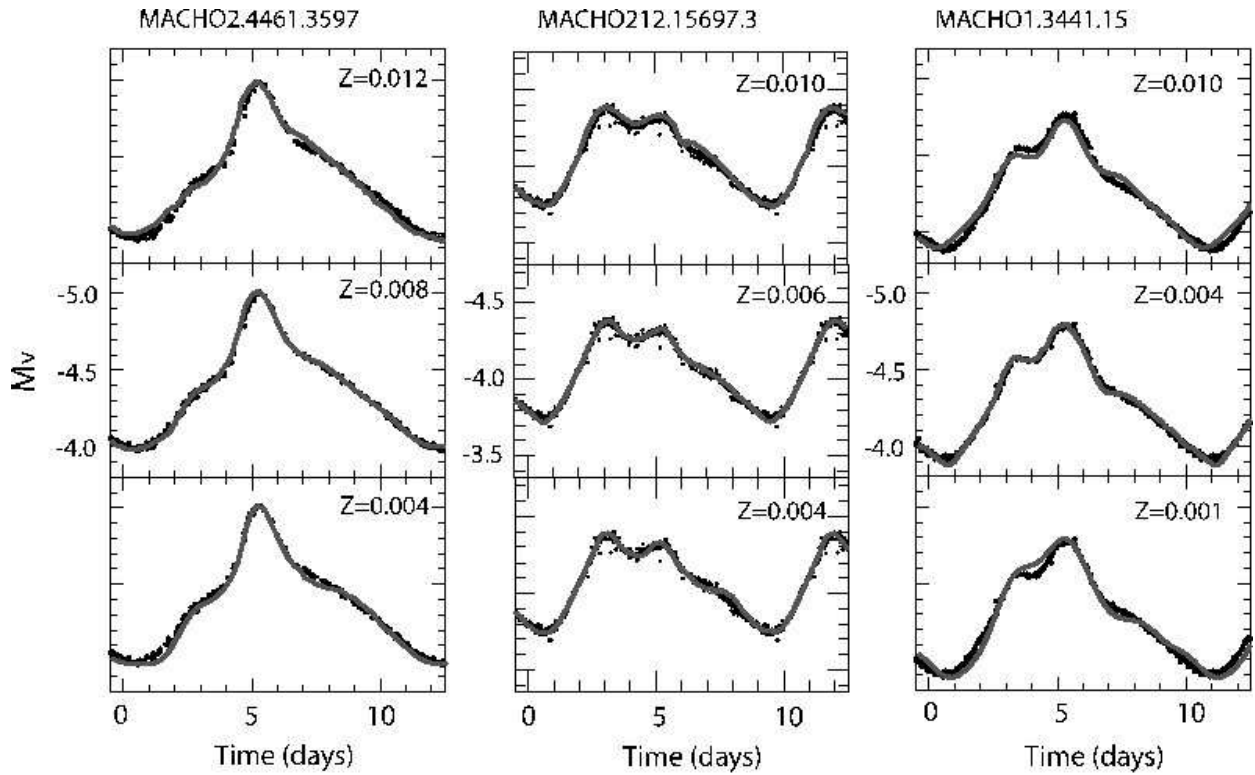}
\caption{The effect of varying metallicity on the model light curve for
  MACHO~2.4661.3597, MACHO~212.15697.3 and MACHO~1.3441.15. Decreasing the
  metallicity shifts the bump at $t=5-8d$ to later phase. The central panel is
  the best fitting model. In the case of MACHO~212.15697.3 this corresponds to
  the central panel of Figure \ref{fig5panel}.\label{figstar2} }
\end{figure}

\begin{figure}
\plotone{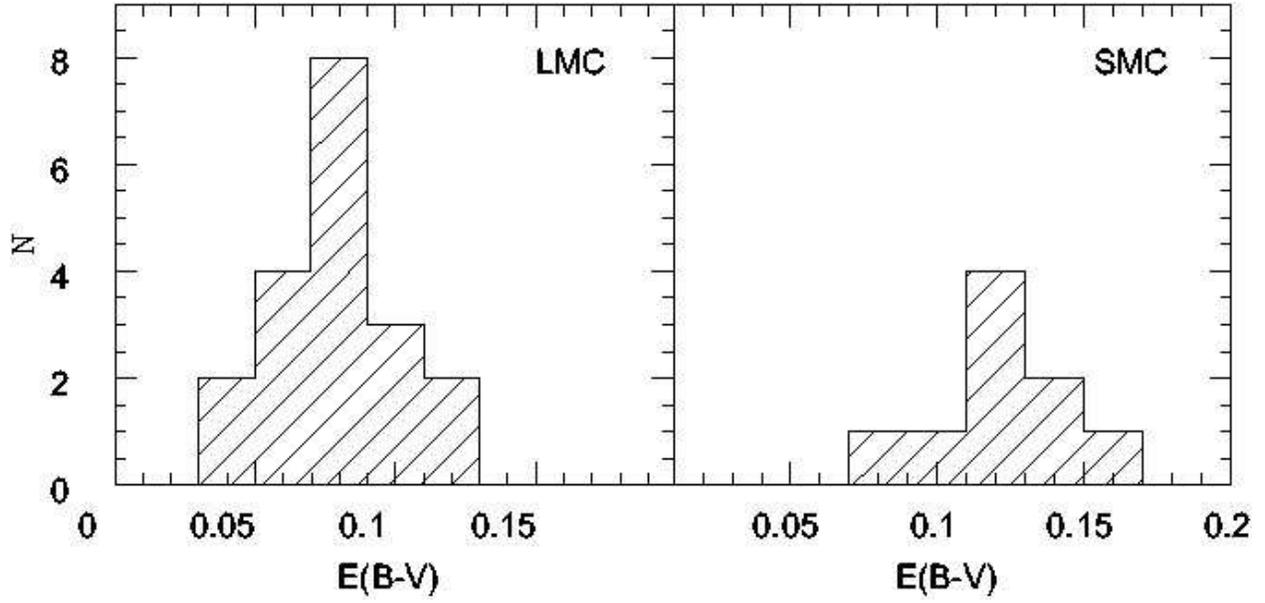}
\caption{Histograms of the derived reddenings of our sample.\label{redfig}}
\end{figure}

\begin{figure}
\plotone{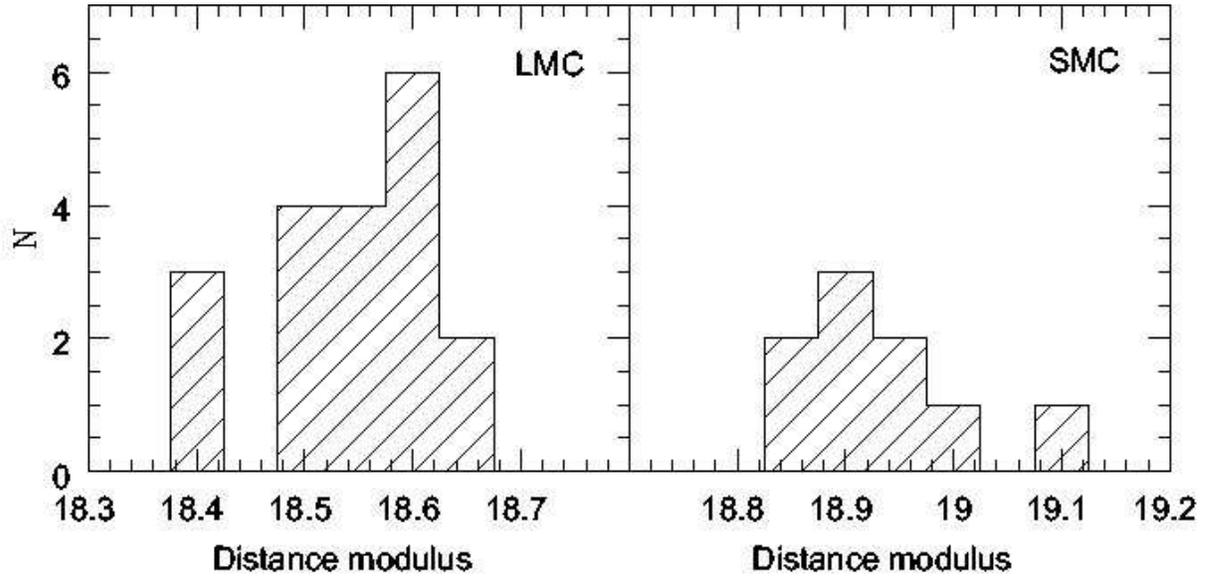}
\caption{Histograms of the derived distance modulus of our sample.\label{dmfig}}
\end{figure}

\begin{figure}
\plotone{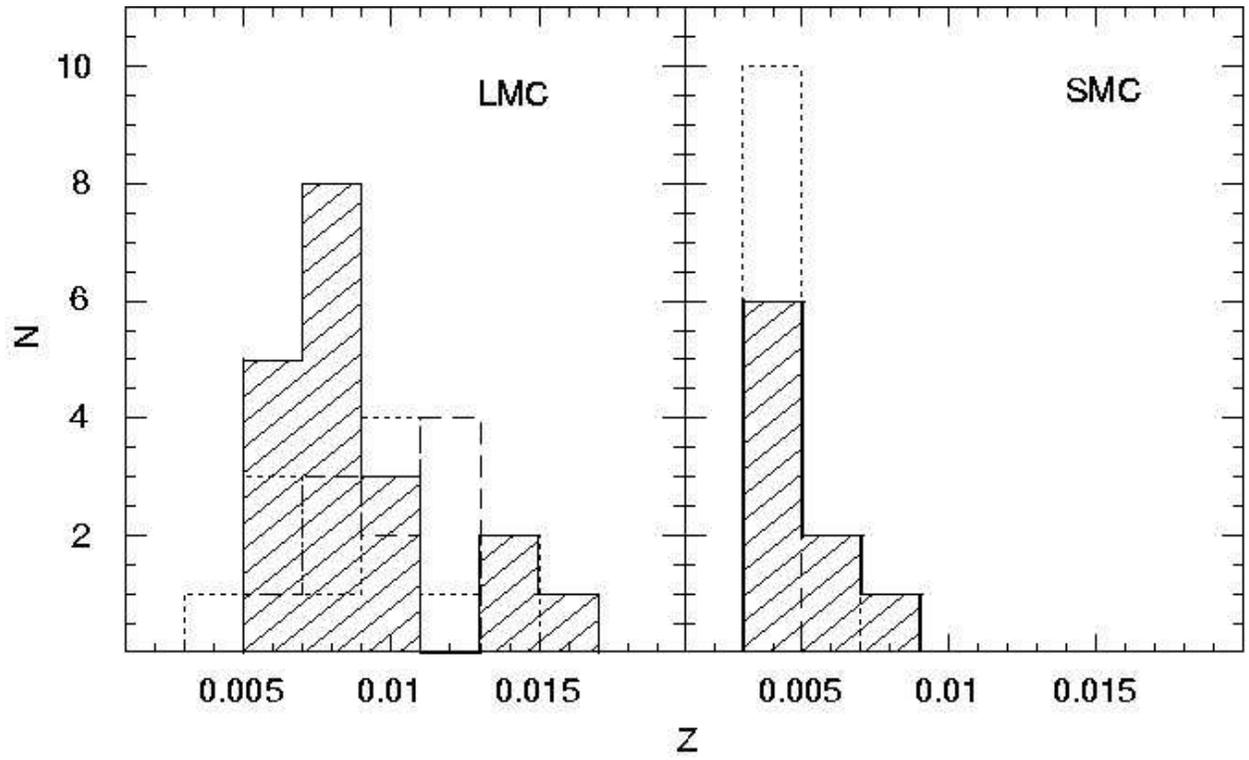}
\caption{Histograms of the derived metallicity of our sample. Overlaid are the histograms from \cite{rom05} (dotted) and \cite{luc98} (dashed).\label{zfig}}
\end{figure}

\begin{figure}
\plotone{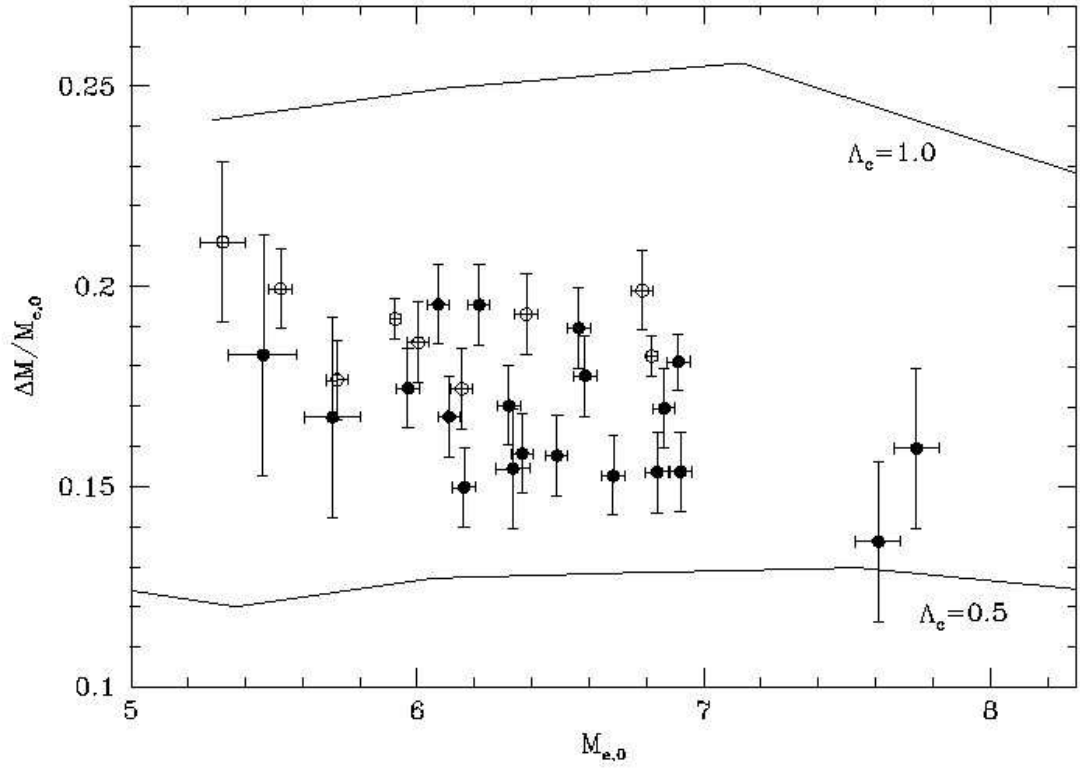}
\caption{The difference $\Delta M = M_{E,0} - M_{P}$ between the pulsation
mass, $M_{P}$, and classical evolutionary mass, $M_{E,0}$, normalized by the
$M_{E,0}$. Open circles are those objects from the SMC. Overlaid are the loci
of models that incorporate different levels of convective core overshoot from
\cite{gir00} and Bressan (2001).}\label{delmfig}
\end{figure}

\begin{figure}
\plotone{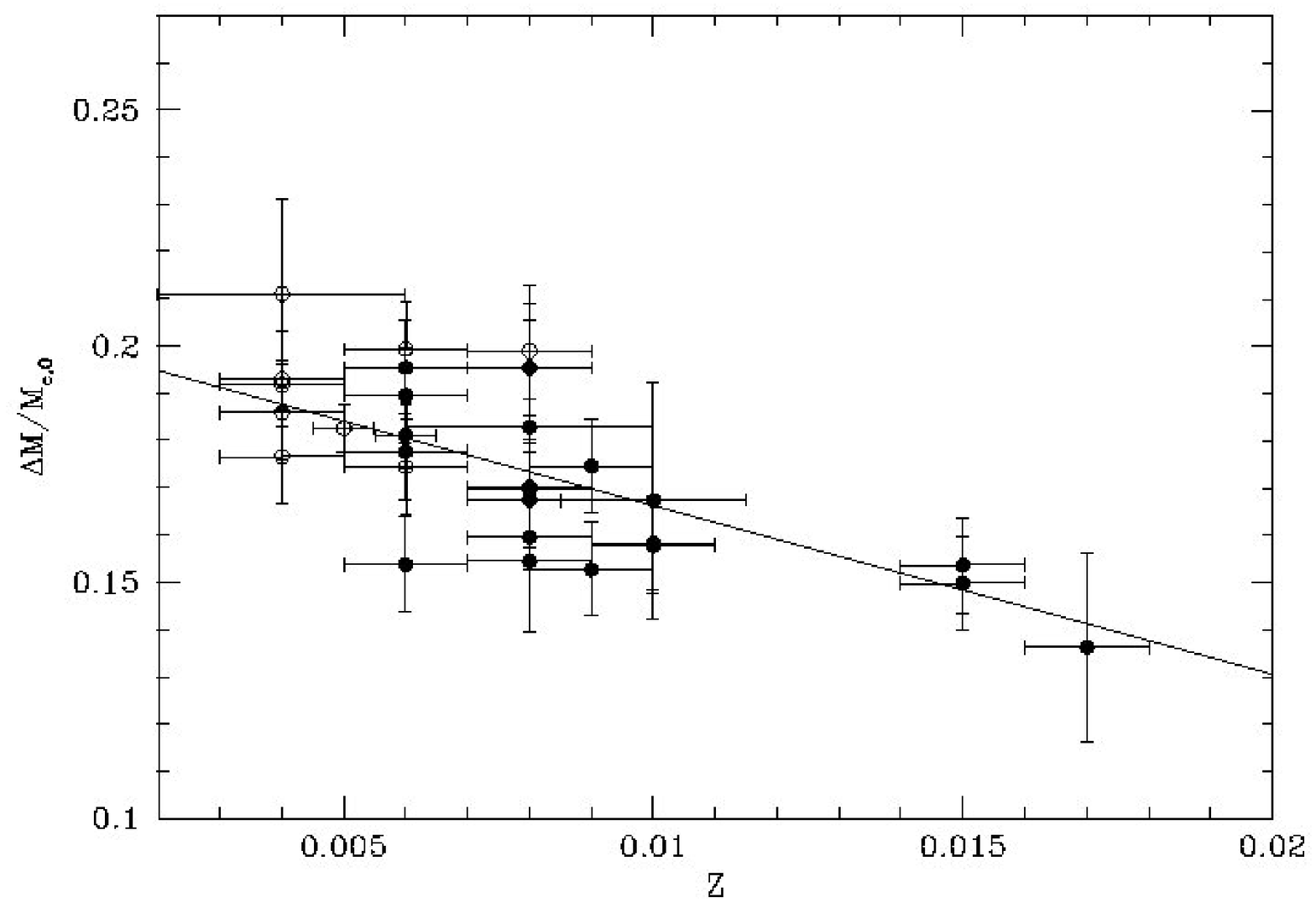}
\caption{The difference $\Delta M = M_{E,0} - M_{P}$ between the pulsation
mass, $M_{P}$, and classical evolutionary mass, $M_{E,0}$, normalized by
$M_{E,0}$ as a function of metallicity. Open circles are those objects from
the SMC. The line is the line of best fit to the data
points.}\label{delm_zfig}
\end{figure}

\end{document}